\documentclass[preprint,showpacs,preprintnumbers,amsmath,amssymb]{revtex4}

\usepackage{graphicx}
\usepackage{dcolumn}
\usepackage{bm}
\usepackage{multirow}
\usepackage{color}
\usepackage{CJK}

\begin{document}

\title{Systematic investigations of positive-parity doublet bands with three-quasiparticle configurations in $^{125,127,129,131}$Cs}

\author{Rui Guo}

\author{Wu-Ji Sun}

\author{Jian Li}
\email{jianli@jlu.edu.cn}

\author{Dong Yang}
\email{dyang@jlu.edu.cn}

\author{Yonghao Liu}

\author{Chengkun Ru}

\author{Jihuai Chi}

\affiliation{College of Physics, Jilin University, Changchun 130012, China}

\begin{abstract}
The experimental features of positive-parity doublet bands in the odd-\emph{A} cesium isotopes $^{125,127,129,131}$Cs, including angular momentum alignment, energy staggering, $B(M1)/B(E2)$ etc. are studied systematically and compared to those of the candidate chiral bands in the adjacent odd-odd Cs isotopes. The configuration assignments and the dynamics of these bands are discussed. The self-consistent tilted axis cranking relativistic mean-field calculations are performed with configuration reassigned to these bands. The experimental level schemes of four nuclei are well reproduced, and the calculations also show four nuclei have obvious triaxial deformations and thus support the candidate chiral doublet bands in $^{125,127,129,131}$Cs.

\end{abstract}

\maketitle

\section{Introduction}
Chirality is a popular symmetry occurring in nature, such as
chemistry, biology and physics. In nuclear physics, chirality arises from the coupling of the angular momenta in a triaxially deformed
nuclei~\cite{Frauendorf1997Nucl.Phys.A131} and the predicted patterns of
spectra exhibiting chirality---chiral doublet bands---were
experimentally observed in 2001~\cite{Starosta2001Phys.Rev.Lett.971}, i.e., the existence of one pair of $\Delta I = 1$ bands (normally regarded as nearly degenerate) with the same parity.
Moreover, two pairs of chiral bands with different configurations in $^{105}$Rh were discussed, respectively, in Refs.~\cite{PhysRevC.69.024317,Timar2004Phys.Lett.B178}. The possible existence of multiple chiral doublet bands in one nucleus was demonstrated by searching for triaxial chiral configurations in Rh isotopes based on constrained relativistic mean-field calculations in Ref.~\cite{Meng2006Phys.Rev.C037303}, which introduced the acronym M$\chi$D for multiple chiral doublet bands.
Up to now, nuclear chirality has attracted a lot of attentions and many chiral candidate nuclei have been reported
experimentally in the $A \sim
80, 100, 130$, and 190 mass regions of the nuclear chart, see, e.g., Refs.~\cite{Starosta2001Phys.Rev.Lett.971,Zhu2003PhysRevLett.91.132501,PhysRevLett.92.032501,JOSHI2004135,PhysRevC.69.024317,Timar2004Phys.Lett.B178,Grodner2006PhysRevLett.97.172501,Joshi2007PhysRevLett.98.102501,Mukhopadhyay2007PhysRevLett.99.172501,Grodner2011Phys.Lett.B46,Wang2011Phys.Lett.B40}, as well as the reviews in Refs.~\cite{Meng2010J.Phys.G064025,Meng2016book,XIONG2019193}. In addition, several multiple doublet candidate nuclei have been proposed or reported in Refs.~\cite{Li2011Phys.Rev.C037301,Qi2013PhysRevC.88.027302,Ayangeakaa2013PhysRevLett.110.172504,Kuti2014PhysRevLett.113.032501,Liu2016PhysRevLett.116.112501}.

In particular, the reported candidate chiral nuclei in $A\sim130$ mass region form a ¡°chiral island¡±~\cite{Starosta2001Phys.Rev.Lett.971}, where the cesium isotopes have the most candidate chiral nuclei.
In all those candidate odd-odd cesium isotopes, i.e., $^{122,124,126,128,130,132}$Cs~\cite{Yong-Num2005J.Phys.GB1,Selvakumar2015PhysRevC.92.064307,Grodner2011Phys.Lett.B46,Grodner2006PhysRevLett.97.172501,130Cs,Rainovski2003PhysRevC68.024318},
one pair of chiral doublet bands is observed. Moreover, those chiral candidates consist of configuration $\pi h_{11/2}\otimes\nu h^{-1}_{11/2}$, i.e., the unpaired odd proton particle lies in the lower part of intruder orbital $h_{11/2}$, whereas
it is in the middle to upper part for the odd neutron.

For the neighboring odd-$A$ cesium isotopes, three-quasiparticle configurations based on high-$j$ proton particle and neutron hole configuration $\pi h_{11/2}\otimes\nu h^{-1}_{11/2}$ can be easily constructed and therefore the nuclear chirality is highly expected. In $^{125, 129, 131}$Cs, a pair of bands has been observed involving the configuration of the $\pi h_{11/2}\otimes\nu h_{11/2}^{-1}$ component with the third neutron quasiparticle in the orbital of ${s_{1/2}}/{d_{3/2}}/{g_{7/2}}$~\cite{125,129,131}.
Moreover, the nearly degenerate doublet bands indicate the possible existence of chirality in $^{125, 129, 131}$Cs, although not discussed specifically and systematically in those Refs.~\cite{125,129,131}. On the other hand, the bands 4 and 6 in $^{127}$Cs ~\cite{127} show very good similarity in level scheme as the doublet bands in $^{125, 129, 131}$Cs,
but assigned a different configuration of $\pi g_{7/2}/d_{5/2}\otimes \nu h^{-2}_{11/2}$.
Based on the above statements, it is very necessary to investigate whether the similar pair of coupled bands in $^{127}$Cs have the same configuration as in $^{125, 129, 131}$Cs.

On the other hand, the adiabatic and configuration-fixed constrained triaxial relativistic mean-field (RMF) approach has been performed to study the triaxial deformation of odd-\emph{A} $^{125,127,129,131}$Cs~\cite{MX125131}. The existence of chirality, especially the M$\chi $D phenomenon, is demonstrated and expected based on different high-$j$ particle-hole configurations and triaxial deformations.
However, the rotation effect is not considered in the above calculations. Based on the covariant density functional theory (CDFT)~\cite{Ring1996Prog.Part.Nucl.Phys.193,Vretenar2005Phys.Rep.101,Meng2006Prog.Part.Nucl.Phys.470,Meng2016book},
which takes into account the Lorentz invariance, the tilted axis
cranking (TAC) calculations have been successfully applied to describe
many collective structural phenomena, such as magnetic rotation~\cite{Peng2008Phys.Rev.C024313,Zhao2011Phys.Lett.B181,Yu2012Phys.Rev.C024318,Li2012Nucl.Phys.A34,Li2013PhysRevC.88.014317,Peng2015Phys.Rev.C044329} and anti-magnetic rotation~\cite{Zhao2011Phys.Rev.Lett.122501,Zhao2012Phys.Rev.C054310,Li2012Phys.Rev.C057305,Peng2015Phys.Rev.C044329,Sun2016ChinPhysC.40.084101}, transition from collective to chiral rotation~\cite{Zhao2015PhysRevC.92.034319}, and rotations with an exotic rod
shape~\cite{PhysRevLett.115.022501}. Thus it is also necessary to check the existence of chirality after considering the rotation effect.

In this paper, the configuration assignments for the doublet bands in $^{125,127,129,131}$Cs are discussed first.
Then the experimental characteristics of these doublet bands are studied systematically, and also compared with the chiral double bands in the neighboring odd-odd nuclei $^{124,126,128,130}$Cs.
At last, self-consistent tilted axis cranking relativistic mean-field (TAC-RMF) theory~\cite{Peng2008Phys.Rev.C024313,Zhao2011Phys.Lett.B181,Meng2013,Zhao2017PhysLettB773} is performed to investigate the triaxial deformation of these bands with assigned configurations.

\section{Systematic investigation of experimental features}
Four pairs of doublet bands with positive parity, i.e., bands 3 and 7 in $^{125}$Cs~\cite{125new}, bands 4 and 6 in $^{127}$Cs~\cite{127}, bands 1 and 10 in $^{129}$Cs~\cite{129}, and bands 2 and 3 in $^{131}$Cs~\cite{131}, are labeled as the band 1 and band 2 here, respectively.
It should be noted that those four pairs of doublet bands show similar structure in energy spectrum. For
example, the energy differences between bands 1 and 2 in those four nuclei are relatively small and almost less than 300 keV.
In addition, the excitation energies of bandhead ($19/2^+$) are all about 2 MeV higher than the bandhead ($11/2^-$) of the one-quasiparticle yrast band with configuration $\pi h_{11/2}$~\cite{127,129,131,125new}.

\begin{figure}[h!]
 \centering
 \includegraphics[width=8.5cm]{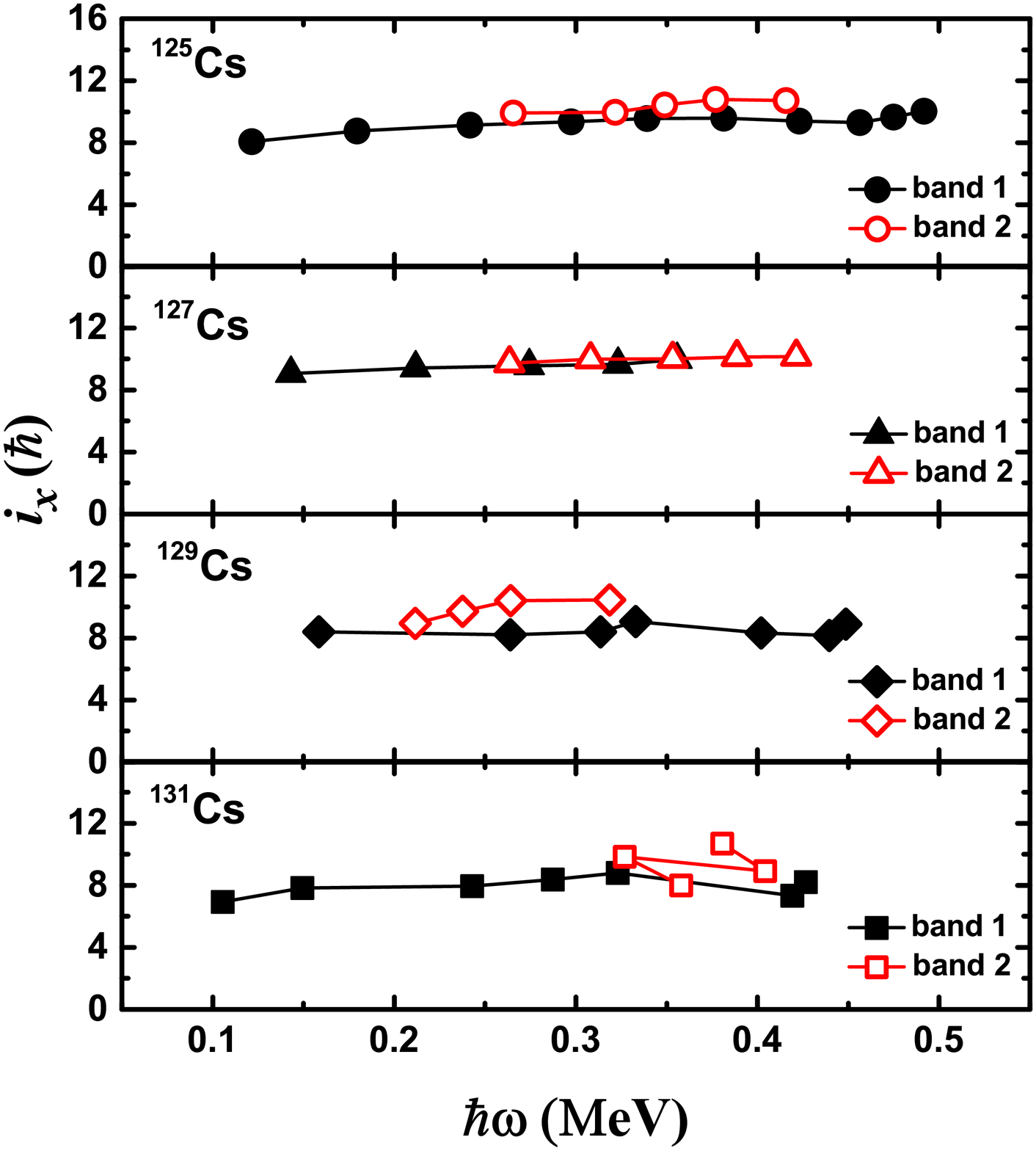}\vspace*{-0.5cm}
  \caption{(Color online) Experimental alignments of doublet bands in $^{125,127,129,131}$Cs. Harris parameters~\cite{J0J1}($J_{0}=5.8\hbar^{2}$MeV$^{-1}$ and
$J_{1}=50.8\hbar^{4}$MeV$^{-3}$) are used to subtract the angular momenta of the core.}
 \label{fig1}
 \end{figure}

To further investigate the similarities of these doublet bands, experimental alignments ($i_{x}$) as a function of rotational frequency $\hbar\omega$ for band 1 and band 2 in $^{125,127,129,131}$Cs are given in Fig.~\ref{fig1}.
The Harris parameters~\cite{J0J1} of $J_{0}=5.8\hbar^{2}$MeV$^{-1}$ and $J_{1}=50.8\hbar^{4}$MeV$^{-3}$ are adopted to subtract the angular momenta of the core.
It could be seen that all these bands show large initial alignments~$\approx 8\hbar$ and then remain constant as frequency increases from 0.1-0.45 MeV and the differences in the alignment values of the doublet bands in each nucleus are very small, which clearly exhibits the similar rotation structure in those four pairs of doublet bands.

 \begin{figure}[h!]
 \centering
 \includegraphics[width=8.5cm]{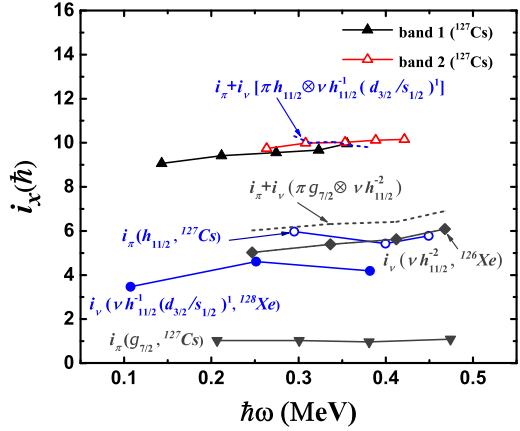}\vspace*{-0.5cm}
  \caption{(Color online) Experimental alignments of doublet bands in $^{127}$Cs, in comparison with the alignments of relevant bands with one-quasiparticle configurations $\pi h_{11/2}$ and $\pi g_{7/2}$ in $^{127}$Cs~\cite{127} and two-quasiparticle configurations $\nu h^{-1}_{11/2}({d_{3/2}}/{s_{1/2}})^1$ and $\nu h_{11/2}^{-2}$ in neighboring even-even nuclei $^{128}$Xe~\cite{Xe128} and $^{126}$Xe~\cite{Xe126}, respectively. The same Harris parameters, i.e., $J_{0}=5.8\hbar^{2}$MeV$^{-1}$ and
$J_{1}=50.8\hbar^{4}$MeV$^{-3}$ are adopted here.}
 \label{fig2}
 \end{figure}
Considering the shell structure in the $A\sim130$ mass region
and the positive parity of the doublet bands,
there are two possible configurations, i.e., $\pi g_{7/2}/d_{5/2}\otimes\nu h_{11/2}^{-2}$ and $\pi h_{11/2}\otimes\nu h^{-1}_{11/2} g_{7/2}/{d_{3/2}}/{s_{1/2}}$. To study and distinguish the two possible configurations for band 1 and band 2, especially in $^{127}$Cs, the alignment additive rule, which has been successfully applied to the configuration assignment of rotational bands in the mass regions of $A\sim130$~\cite{PhysRevC.58.1849} and $A\sim160$~\cite{DRISSI1986313}, is adopted.
In Fig.~\ref{fig2}, experimental alignments of doublet bands in $^{127}$Cs are compared with the alignments of relevant bands with one-quasi-proton configurations $\pi h_{11/2}$ and $\pi g_{7/2}$ in $^{127}$Cs~\cite{127}, two-quasi-neutron configuration $\nu h^{-1}_{11/2}({d_{3/2}}/{s_{1/2}})^1$ in neighboring even-even nuclei $^{128}$Xe~\cite{Xe128} and a pair of neutron $h_{11/2}$ quasiparticles i.e., $\nu h_{11/2}^{-2}$ in $^{126}$Xe~\cite{Xe126}.
The same Harris parameters as used in Fig.~\ref{fig1} are adopted here to obtain the experimental alignments.
It can be easily seen that the final alignment values of three-quasiparticle configuration $\pi h_{11/2}\otimes\nu h^{-1}_{11/2} ({d_{3/2}}/{s_{1/2}})^1$ ($\approx 9.5\hbar$), obtained from the sum of two quasi-particle configuration $\nu h_{11/2}^{-1}({d_{3/2}}/{s_{1/2}})^1$ in $^{128}$Xe ($\approx 4\hbar$) and one quasi-particle configuration $\pi h_{11/2}$ in $^{127}$Cs ($\approx 5.5\hbar$), is in good agreement with the experimental alignments of the doublet bands.
On the contrary, the alignments for $\pi g_{7/2}\otimes\nu h_{11/2}^{-2}$ ($\approx 6\hbar$), obtained from the sum of two quasi-particle configuration $\nu h_{11/2}^{-2}$ in $^{126}$Xe ($\approx 5\hbar$) and one quasi-particle configuration $\pi g_{7/2}$ in $^{127}$Cs ($\approx 1\hbar$), deviate much from the data~\cite{127}.
Thus, the possible configuration for band 1 and 2 in $^{127}$Cs is $\pi h_{11/2}\otimes\nu h^{-1}_{11/2} g_{7/2}/{d_{3/2}}/{s_{1/2}}$ rather than $\pi g_{7/2}\otimes\nu h_{11/2}^{-2}$.
It should be noted that in $^{127}$Cs, the $\pi g_{7/2}$ band and $\pi d_{5/2}$ band have the similar alignments, and the possible configuration $\pi d_{5/2}\otimes\nu h_{11/2}^{-2}$ is also excluded. Additionally, $^{125}$Cs, $^{129}$Cs, and $^{131}$Cs have the similar alignments and hence are assigned similar configuration.

\begin{figure}[ht]
 \centering
 \includegraphics[width=8.5cm]{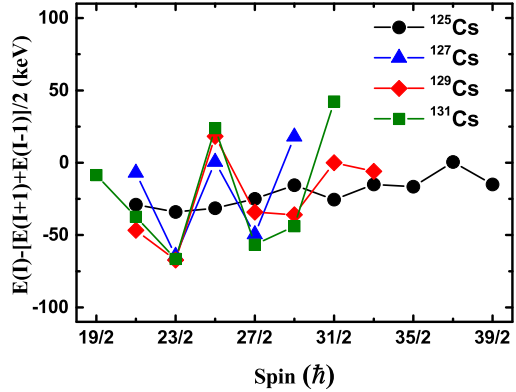}\vspace*{-0.5cm}
  \caption{(Color online) Illustration of signature splitting $E(I)-[E(I+1)+E(I-1)]/2$ vs. spin for the band 1 in $^{125,127,129,131}$Cs.}
 \label{fig3}
 \end{figure}

At present, the three-quasiparticle configurations are involving the high-$j$ shape-driving $\pi h_{11/2}$ and $\nu h_{11/2}$ orbitals as well as a positive-parity quasineutron from the $s_{1/2}$, $d_{3/2}$, or $g_{7/2}$ orbital with a low alignment contribution.
To examine the occupation orbital of the observing neutron, the energy signature splitting, i.e., $E(I)-[E(I+1)+E(I-1)]/2$, as a function of the spin for the band 1 in $^{125,127,129,131}$Cs is shown in Fig.~\ref{fig3}.
It is easy to see that band 1 of $^{125}$Cs exhibits near to zero signature splitting, while relatively large signature splitting is observed for $^{127,129,131}$Cs. It has already been pointed out in Ref.~\cite{125} that to understand the near to zero signature splitting,
the last neutron in $^{125}$Cs should be occupying the high-$\Omega$ orbital at the top of $g_{7/2}$ shell, and the configuration $\pi h_{11/2}\otimes\nu h^{-1}_{11/2} g^{-1}_{7/2}$ is favoured for the doublet bands in $^{125}$Cs. On the other hand, the obvious signature splitting favours the involvement of the $s_{1/2}$ or $d_{3/2}$ orbital in $^{127,129,131}$Cs,
and therefore, the configuration $\pi h_{11/2}\otimes\nu h^{-1}_{11/2} ({d_{3/2}}/{s_{1/2}})^1$ is favoured for the doublet bands, which are different from the previous configuration assignments, i.e.,
$\pi g_{7/2}(d_{5/2})\otimes\nu h_{11/2}^{-2}$ in $^{127}$Cs~\cite{127} and $\pi h_{11/2}\otimes\nu h_{11/2}^{-1}g_{7/2}$ in $^{129}$Cs~\cite{129}.

\begin{figure}[h]
 \centering
 \includegraphics[width=10cm]{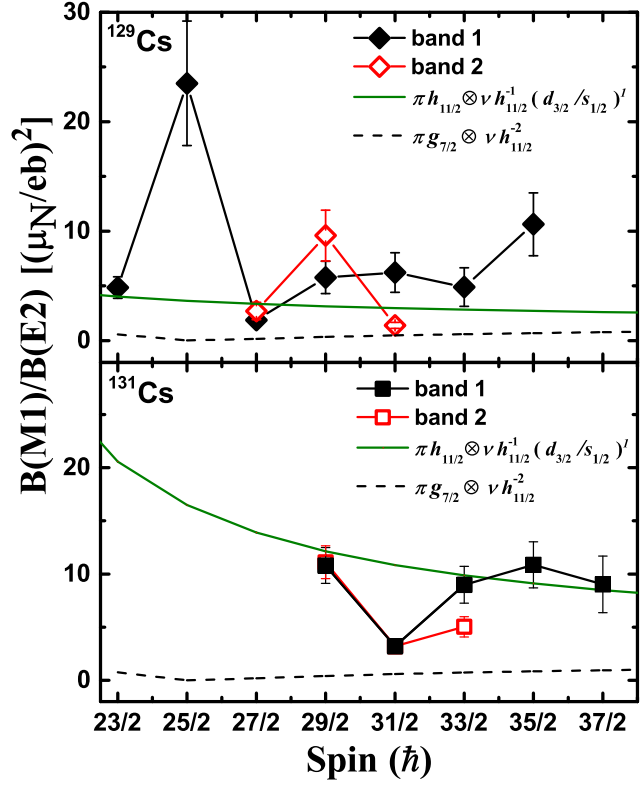}\vspace*{-0.5cm}
  \caption{(Color online) The reduced transition probabilities $B(M1)/B(E2)$ values as a function of the spin in the calculations compared with the experimental data~\cite{129,131} of band 1 and band 2 in $^{129}$Cs and $^{131}$Cs, respectively. Parameters used in the calculations of the predicted values: Q$_{0}$ = 2.98 eb ($^{129}$Cs), Q$_{0}$ = 2.5 eb ($^{131}$Cs), $g_{R} = 0.43 (^{129}$Cs), $g_{R} = 0.42 (^{132}$Cs), $g_{p}$($h_{11/2}$) = 1.17, $g_{p}(g_{7/2}) = 0.72, g_{n}(h_{11/2})=0.343, g_{n}(d_{3/2}/s_{1/2}) =0.459, i_{p}(h_{11/2}) = 5, i_{p} (g_{7/2}) = 1.5, i_{n}(h_{11/2}) = 1.5     $, and $  i_{n}(d_{3/2}/s_{1/2}$) = 0.5.}
 \label{fig4}
 \end{figure}

To further check the configuration assignments of the doublet bands, the ratios of reduced transition probabilities $B(M1)/B(E2)$ calculated with the geometrical model~\cite{BM1BE2} in $^{129}$Cs and $^{131}$Cs as a function of the spin are compared with the corresponding experimental data extracted from Refs.~\cite{129,131} in Fig.~\ref{fig4}.
It is easy to see that the experimental $B(M1)/B(E2)$ ratios for the intraband transitions of the two bands are very similar in experimental errors, which indicates that the doublet bands are based on similar configurations in $^{129}$Cs and $^{131}$Cs, respectively.
On the other hand, the calculated $B(M1)/B(E2)$ ratios in $^{129}$Cs and $^{131}$Cs with configuration $\pi h_{11/2}\otimes\nu h^{-1}_{11/2} ({d_{3/2}}/{s_{1/2}})^1$ are more close to the experimental ratio, while the corresponding calculated ratios with configuration $\pi g_{7/2}\otimes\nu h_{11/2}^{-2}$ deviate from the experimental value. These further support the present configuration assignment.

\begin{figure}[h!]
 \centering
 \includegraphics[width=15cm]{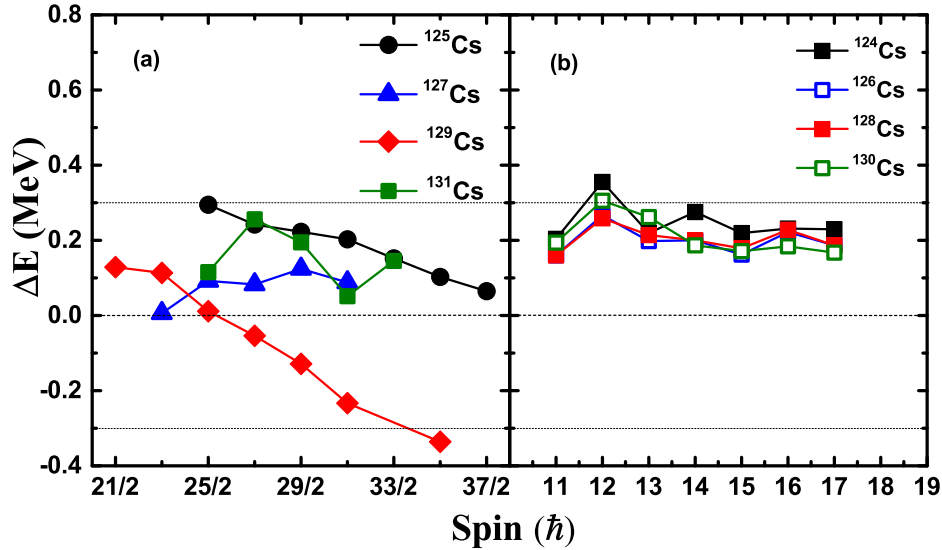}\vspace*{-0.5cm}
  \caption{(Color online) Comparison of energy differences $\triangle{E}$, i.e., $E(I)_{Band1}-E(I)_{Band2}$ as a function of the spin of doublet bands in $^{124,126,128,130}$Cs and $^{125,127,129,131}$Cs, respectively.}
 \label{fig5}
 \end{figure}

The present configuration assignments of high-$j$ proton particle and neutron hole configurations for the doublet bands in $^{125,127,129,131}$Cs are favorable for the chirality, which deserves more investigations. In Fig.~\ref{fig5}, the energy differences for doublet bands, i.e., $\triangle{E}=E(I)_{Band1}-E(I)_{Band2}$, in $^{125,127,129,131}$Cs and neighboring odd-odd nuclei $^{124,126,128,130}$Cs are shown. It should be noted that the energy differences of doublet bands in $^{125,129,131}$Cs are almost less than 300 keV and $\triangle{E}$ in $^{127}$Cs is even below 100 keV. Thus the behavior of almost degenerate $\triangle{I} = 1$ doublet bands is clearly exhibited, similar as in $^{124,126,128,130}$Cs.
Furthermore, the energy differences $\triangle{E}$ for the doublet bands in $^{125}$Cs decrease gradually with the increasing spin, i.e., from 300-50 keV, which may be related to the transition of chiral vibration and static chirality~\cite{Starosta2001Phys.Rev.Lett.971}.
In particular, the $\triangle{E}(I)$ in $^{129}$Cs decreases and becomes negative with the increasing spin, i.e., the band 1 is crossed by band 2 after $I=12.5$. This phenomenon has also been observed for the chiral doublet bands in $^{112}$Ru~\cite{PhysicsLettersB670307}, reflecting a dynamical character of chirality.

\begin{figure}[h!]
 \centering
 \includegraphics[width=15cm]{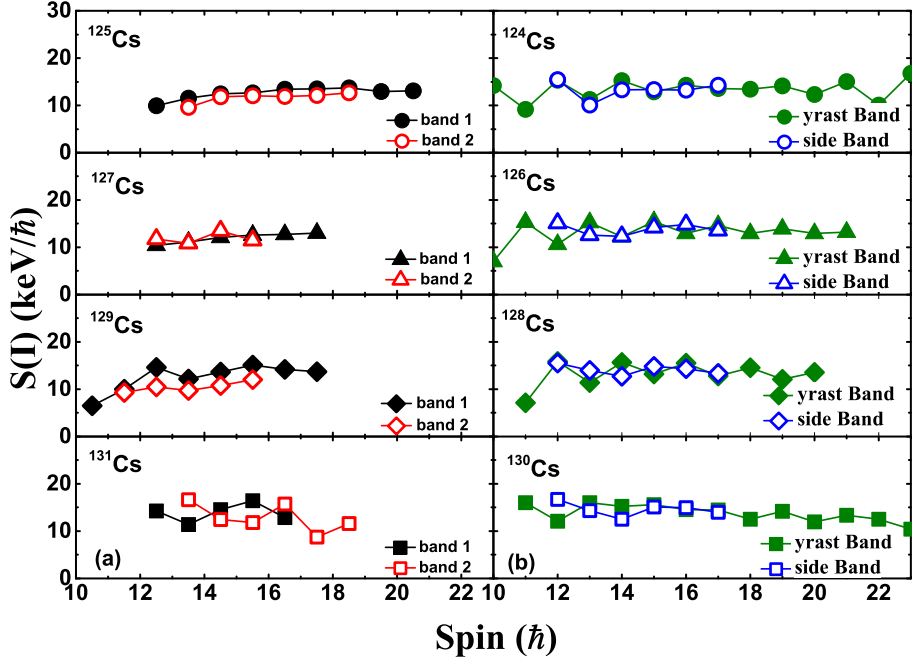}\vspace*{-0.5cm}
  \caption{(Color online) $S(I)$ as a function of the spin for the doublet bands in $^{125,127,129,131}$Cs (left panel)~\cite{125new,127,129,131} compared to those in $^{124,126,128,130}$Cs (right  panel)~\cite{Selvakumar2015PhysRevC.92.064307,Grodner2011Phys.Lett.B46,Grodner2006PhysRevLett.97.172501,130Cs}, respectively.}
 \label{fig6}
 \end{figure}

In addition, the energy staggering parameter $S(I)=[E(I)-E(I-1)]/2I$ is also an important criterion for chirality since first proposed and discussed in Ref.~\cite{PhysRevLett.92.032501}.
It should possess a smooth independence of spin since the angular momenta of proton particle and neutron hole orbital are both perpendicular
to the core rotation in the chiral geometry~\cite{PhysRevLett.92.032501}.
In Fig.~\ref{fig6}, the $S(I)$ for the doublet bands in $^{125-131}$Cs is given and compared with those in $^{124,126,128,130}$Cs.
It can be seen that $S(I)$ is almost constant with increasing spin or shows very small staggering around 10 keV/$\hbar$. Furthermore, the $S(I)$ of band 1 and band 2 in $^{125-131}$Cs is very close to each other. Thus the two characters of $S(I)$ support the nuclear chirality for the doublet bands.

Chiral doublet bands should have similar $B(M1)/B(E2)$ values and staggering behavior.
In Fig.~\ref{fig4}, the experimental $B(M1)/B(E2)$ ratios of the doublet bands, only available in $^{129,131}$Cs, have been given.
It is easy to see that the $B(M1)/B(E2)$ values within experimental errors for each state
in $^{129,131}$Cs are in reasonable agreement, and the values also exhibit a clear staggering pattern with increasing spin. Moreover, by comparing $^{129,131}$Cs with the adjacent odd-odd nuclei $^{128,130}$Cs~\cite{Grodner2006PhysRevLett.97.172501,130Cs}, it can be found that the $B(M1)/B(E2)$ values are similar, which indicates the doublet bands in odd-\emph{A} Cs isotopes have similar chiral geometry required for chiral doublets.

\section{Theoretical calculation}

In the following, the structure of rotational bands in $^{125,127,129,131}$Cs are investigated by tilted axis cranking relativistic mean-field (TAC-RMF) approach. In contrast to its non-relativistic counterparts~\cite{RevModPhys.75.121}, the CDFT, including relativistic mean-field (RMF) framework with point coupling or mesonic exchange interaction~\cite{Serot1986Adv.Nucl.Phys.1,Nikolaus1992Phys.Rev.C46,WenHuiLongPhysicsLettersB},
takes the fundamental Lorentz symmetry into account from the very beginning so that naturally takes care of the important spin degree of freedom,
resulting in great successes on many nuclear phenomena~\cite{Meng2016book,RevModPhys.75.121,Ring1996Prog.Part.Nucl.Phys.193,Vretenar2005Phys.Rep.101,Meng2006Prog.Part.Nucl.Phys.470}.
Moreover, without any additional parameters, the rotation excitations can be described self-consistently with the tilted axis cranking relativistic mean-field approach~\cite{Meng2013,Meng2016book}.

In the TAC-RMF theory, nuclei are characterized by the relativistic fields $S(\bm{r})$ and $V^{\mu}(\bm{r})$ in the Dirac equation in the rotating frame with a constant angular velocity vector $\bm{\Omega}$ as
\begin{equation}\label{DiracEq}
 [\bm{\alpha}\cdot(-i\nabla- \bm{V} ) +\beta (m + S)+V - \bm{\Omega}\cdot \bm{\hat{J}}]\psi_i =\varepsilon_i\psi_i,
\end{equation}
where $\bm{\hat{J}}=\bm{\hat{L}}+\frac{1}{2}\bm{\hat{\Sigma}}$ is the total angular momentum of the nucleon spinors, and $\varepsilon_i$ represents the single-particle
Routhians for nucleons. The detailed formalism and numerical techniques can be seen in Refs.~\cite{Peng2008Phys.Rev.C024313,Zhao2011Phys.Lett.B181,Zhao2012Phys.Rev.C054310}. A spherical harmonic oscillator basis with ten major shells is adopted to solve the Dirac equation. The point-coupling interaction PC-PK1~\cite{Zhao2010Phys.Rev.C054319} is used for the Lagrangian, while the pairing
correlations are neglected.

\begin{figure}[h!]
 \centering
 \includegraphics[width=8.5cm]{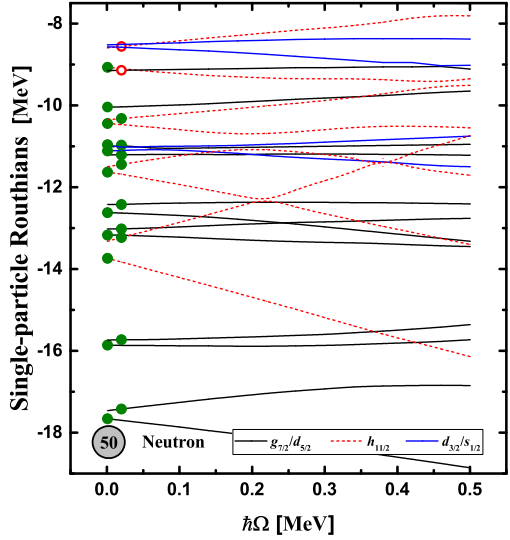}\vspace*{-0.5cm}
  \caption{(Color online) Single-neutron Routhians near the Fermi surface in $^{125}$Cs as a function of the rotational frequency for configuration $\pi h_{11/2}\otimes\nu h^{-1}_{11/2} {g_{7/2}^{-1}}$. The single neutron levels belonging to the configuration $\nu{g_{7/2}}/{d_{5/2}}$ and $\nu{d_{3/2}}/{s_{1/2}}$ are marked by solid black and blue lines, respectively, and the levels belonging to $\nu h_{11/2}$ with negative parity are indicated by dashed red lines.
  The green filled circles indicate the occupied levels and the open red circles indicate the holes in the neutron $h_{11/2}$ and ${g_{7/2}}/{d_{5/2}}$ shell.}
 \label{fig7}
 \end{figure}

The TAC-RMF calculations are based on the above assigned configurations, i.e., $\pi h_{11/2}\otimes\nu h^{-1}_{11/2} {g_{7/2}^{-1}}$ for $^{125}$Cs and $\pi h_{11/2}\otimes\nu h^{-1}_{11/2} ({d_{3/2}}/{s_{1/2}})^1$ for $^{127,129,131}$Cs.
It should be noted that there is a mixing between the low-$j$ orbital $g_{7/2}$ and $d_{5/2}$ in the TAC-RMF calculations, and they can not clearly be distinguished. Taking $^{125}$Cs as an example, the single-neutron Routhians near the Fermi surface in $^{125}$Cs as a function of the rotational frequency for configuration $\pi h_{11/2}\otimes\nu h^{-1}_{11/2} {g_{7/2}^{-1}}$ are shown in Fig.~\ref{fig7}. The single-particle levels with positive (negative) parity are marked by solid (dashed) lines. The green filled circles indicate the occupied levels and the open red circles represent the holes in the neutron $h_{11/2}$ and ${g_{7/2}}/{d_{5/2}}$ shell.

In Fig.~\ref{fig7}, the rotational frequency ($\hbar\Omega$) varies from 0-0.5 MeV. In principle, broken time-reversal symmetry by the cranking field will be recovered without cranking field at $\Omega = 0$, i.e., the levels will be degenerated again as already shown in Fig.~\ref{fig7}.
There are 20 neutrons above the $N = 50$ shell, and it is easy to see that the last two unpaired neutrons are kept fixed in the upper part of $h_{11/2}$ and ${g_{7/2}}/{d_{5/2}}$ shells, respectively. Comparing with $^{125}$Cs, the adding neutrons in $^{127,129,131}$Cs will occupy in the ${d_{3/2}}/{s_{1/2}}$ and $h_{11/2}$ shell, and the corresponding configuration is $\pi h_{11/2}\otimes\nu h^{-1}_{11/2} ({d_{3/2}}/{s_{1/2}})^1$.

\begin{figure}[h!]
 \centering
 \includegraphics[width=15cm]{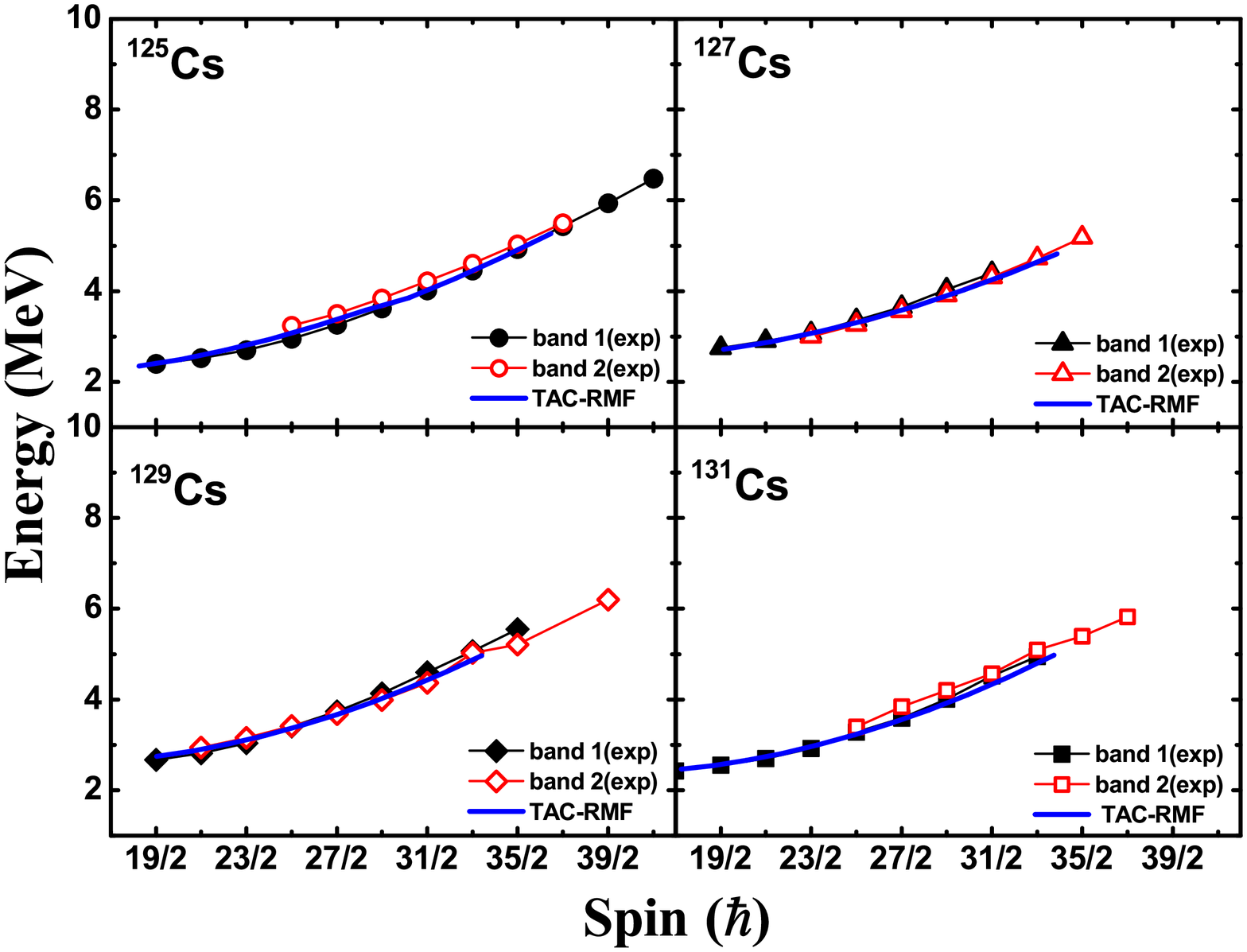}\vspace*{-0.5cm}
  \caption{(Color online) The energy spectra obtained from the TAC-RMF calculations in comparison with the data~\cite{125new,127,129,131} for bands 1 and 2 in $^{125,127,129,131}$Cs, respectively.}
 \label{fig8}
 \end{figure}

In Fig.~\ref{fig8}, the calculated energy spectra in TAC-RMF theory based on corresponding configurations for $^{125,127,129,131}$Cs are shown, and also compared with experimental data of bands 1 and 2~\cite{125new,127,129,131}. It is obvious to see that the experimental energy spectrum is well reproduced by the present self-consistent TAC-RMF calculations, for the band 1. One should note that only the yrast state (line) for a certain configuration can be obtained by TAC-RMF calculations due to the variational principle.

\begin{figure}[h!]
 \centering
 \includegraphics[width=15cm]{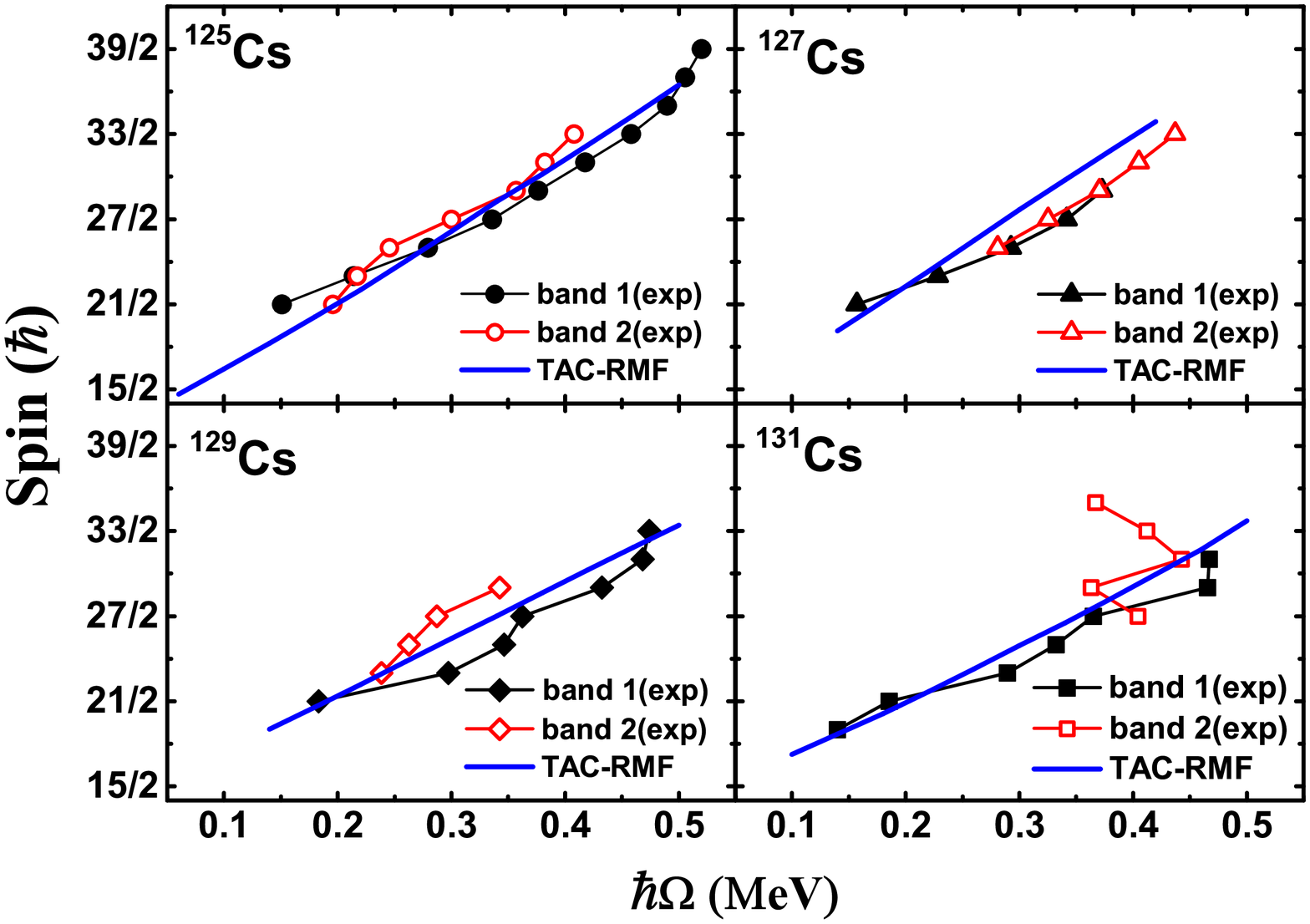}\vspace*{-0.5cm}
  \caption{(Color online) Angular momenta as functions of the rotational frequencies in the TAC-RMF calculations compared with the data~\cite{125new,127,129,131} for bands 1 and 2 in $^{125,127,129,131}$Cs, respectively.}
 \label{fig9}
 \end{figure}

In Fig.~\ref{fig9}, the calculated total angular momenta $I$ with corresponding configuration for $^{125,127,129,131}$Cs determined by the cranking condition $J=\sqrt{I(I+1)}$ are shown as a function of the rotational frequency in comparison with the data. The experimental rotational frequency can be extracted as in Ref.~\cite{Frauendorf1996Z.Phys.A263}, i.e.,
$\hbar\Omega_{exp}=\frac{1}{2}[E_\gamma(I+1\rightarrow I)+E_\gamma(I\rightarrow I-1)]$.
In general, the TAC-RMF results are in good agreement with the band 1 in $^{125,127,129,131}$Cs, which supports the configuration assignments, i.e., $\pi h_{11/2}\otimes\nu h^{-1}_{11/2} {g_{7/2}^{-1}}$ and $\pi h_{11/2}\otimes\nu h^{-1}_{11/2} ({d_{3/2}}/{s_{1/2}})^1$ for the doublet bands in $^{125}$Cs and $^{127,129,131}$Cs, respectively.

\begin{figure}[h!]
 \centering
 \includegraphics[width=8.5cm]{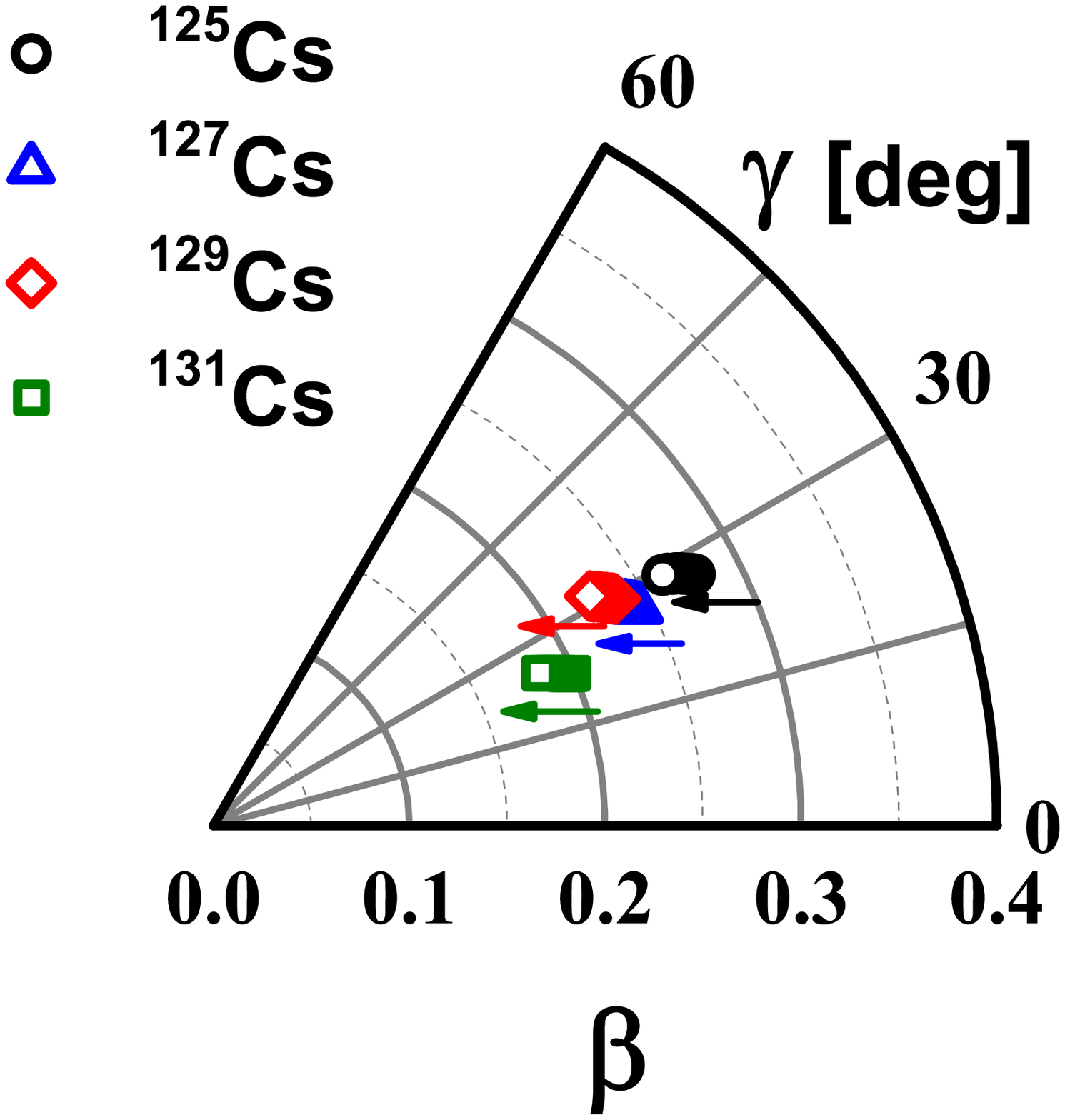}\vspace*{-0.5cm}
  \caption{(Color online) The evolutions of the deformation parameters $\beta$ and $\gamma$ in the TAC-RMF calculations for $^{125}$Cs and $^{127,129,131}$Cs with the configuration $\pi h_{11/2}\otimes\nu h^{-1}_{11/2} {g_{7/2}^{-1}}$ and $\pi h_{11/2}\otimes\nu h^{-1}_{11/2} ({d_{3/2}}/{s_{1/2}})^1$, respectively. }
 \label{fig10}
 \end{figure}

The calculated quadrupole ($\beta$) and triaxial ($\gamma$) deformation parameters by TAC-RMF theory for $^{125,127,129,131}$Cs of assigned configuration are given in Fig.~\ref{fig10} for checking the chirality with rotational mean field.
It could be found that the deformations for configurations $\pi h_{11/2}\otimes\nu h^{-1}_{11/2} {g_{7/2}^{-1}}$ in $^{125}$Cs and $\pi h_{11/2}\otimes\nu h^{-1}_{11/2} ({d_{3/2}}/{s_{1/2}})^1$ in $^{127,129,131}$Cs behave in a similar way, i.e., the triaxial deformations $\gamma$ of the four nuclei is stable and there is only a slight change with the increasing rotational frequency.
For example, for $^{125,127,131}$Cs, the $\gamma$ stays from $27^{\circ}$ to $31^{\circ}$, and $^{129}$Cs keeps between $23^{\circ}$ and $25^{\circ}$. Moreover, the present stable $\beta$ and triaxial deformations support the character of collective motion, in accordance with appearance of chiral rotation. In comparison, the deformation varies much in the magnetic rotation, and the triaxial deformation change can be as large as $30^{\circ}$ for $M1$ bands in $^{60}$Ni~\cite{Zhao2011Phys.Lett.B181}.
In addition, comparing the triaxial deformation in RMF approach without rotation in Ref.~\cite{MX125131}, it is found that considering or without considering the rotational mean field, i.e., the TAC-RMF and RMF calculations both support the existence of chirality in $^{125,127,129,131}$Cs.

It should be noted that the TAC-RMF result here was based on a two-dimensional tilted axis cranking (2DTAC) calculation, which was applied to examine the deformation for the possible configuration of the candidate doublet bands. Theory wise, chiral doublet bands were first investigated in the tilted axis cranking approximation~\cite{Frauendorf1997Nucl.Phys.A131}. Later on, realistic three-dimensional TAC (3DTAC)
approaches, such as 3DTAC based on a Woods¨CSaxon potential combined with the Shell correction method~\cite{PhysRevLett.84.5732} or a Skyrme¨CHartree¨CFock mean field~\cite{Olbratowski2004Phys.Rev.Lett.052501} as well as a relativistic mean field approach~\cite{Zhao2017PhysLettB773}, have been developed to find the chiral solution. In order to describe the nuclear chiral rotation further, the three-dimensional tilted axis cranking is expected to be applied to $^{125,127,129,131}$Cs in future.

\section{Summary}
In conclusion, a systematic study of the doublet bands in $^{125,127,129,131}$Cs has been performed. Based on the analysis of the experimental characteristics and the alignment additive rule etc., the configurations of the doublet bands in these nuclei have been reassigned. The configurations $\pi h_{11/2}\otimes\nu h^{-1}_{11/2} {g_{7/2}^{-1}}$ and $\pi h_{11/2}\otimes\nu h^{-1}_{11/2} ({d_{3/2}}/{s_{1/2}})^1$ are suggested to be favoured for the doublet bands in $^{125}$Cs and $^{127,129,131}$Cs, respectively.
In comparison with the adjacent odd-odd nuclei $^{124-130}$Cs, the experimental characteristics of these nearly degenerate bands are discussed. The investigation finds that these experimental features meet the requirements of the chiral double band, including similar alignments values, and energy staggering parameter $S(I)$ changing gently with increasing spins. In addition, the $B(M1)/B(E2)$ ratios, which are currently only available in $^{129,131}$Cs, also fulfill the experimental feature of the chiral bands.
In addition, the self-consistent tilted axis cranking relativistic mean-field calculations have been performed based on the reassigned configurations for these bands in $^{125,127,129,131}$Cs. The corresponding experimental energy spectra and the angular momenta have been well reproduced, which supports the configuration assignments. The calculated deformation parameters also support the existence of chiral doublet bands in these nuclei.

\begin{acknowledgments}
We are grateful to Professor Y. J. Ma for useful discussions. This work is supported by the National
Natural Science Foundation of China under Grant Nos. 11675063, 11205068, 11475072 and 11847310.
\end{acknowledgments}


\end{document}